\documentclass[aps,prl,preprint]{revtex4-1}
\usepackage[T1]{fontenc}
\usepackage[latin9]{inputenc}
\usepackage{xcolor}
\usepackage{float}
\usepackage{bbding}
\usepackage{amsmath}
\usepackage{amssymb}
\usepackage{graphicx}
\usepackage{nicefrac}

\makeatletter
\usepackage{cleveref}

\providecommand{\newline}{\a}
\makeatother

\usepackage{babel}
\begin{document}


\title{Absence of Walker breakdown in the dynamics of chiral N\'eel domain walls driven by in-plane strain gradients\\
}

\def\correspondingauthor{\footnote{mfa@usal.es}}
\author{ Mouad Fattouhi \correspondingauthor{}, Felipe Garcia-Sanchez, Rocio Yanes, Victor Raposo, Eduardo Martinez and Luis Lopez-Diaz}
\affiliation{Departamento de Fisica Aplicada, Universidad de Salamanca, 37008, Salamanca, Spain}

\date{\today}

\begin{abstract}
We investigate theoretically the motion of chiral N\'eel domain walls in perpendicularly magnetized systems driven by in-plane strain gradients. We show that such strain drives domain walls efficiently towards increasing tensile (compressive) strain for positive (negative) magnetostrictive materials. During their motion a local damping torque that opposes the precessional torque due to the strain gradient arises. This torque prevents the onset of turbulent dynamics, and steady domain wall motion with constant velocity is asymptotically reached for any arbitrary large strain gradient. Withal, velocities in the range of 500 m/s can be obtained using voltage-induced strain under realistic conditions.
\end{abstract}

\maketitle

Reliable, fast and efficient domain wall (DW) motion in perpendicularly magnetized media is a key aspect in the development of new spintronic devices for a variety of applications, such as memory \cite{Parkin_2015}, sensing \cite{Zhang2018}, logic \cite{Franken2012,Luo2020} or neuromorphic computing \cite{Alamdar2021,Yue2019}. It is well known that, when driven by an external force, such as an out-of-plane field, the DW changes its internal structure due to the precessional component of the driving force, which rotates it away from its orientation at rest \cite{hubert2008}. For low values of the driving force a terminal DW angle is reached for which this precessional torque is counterbalanced by the restoring torque that tries to bring it back to its equilibrium orientation, leading to the DW moving rigidly at a constant velocity. Above a certain threshold value, however, this balance is no longer possible and continuous internal DW precession takes place during its motion with the consequent reduction in speed. This so-called Walker breakdown (WB) \cite{Glathe2008} is quite a general phenomenon and it is present, for example, when the driving force is an out-of-plane field \cite{Mougin2007} or a spin polarized current via the spin-transfer torque \cite{Ryu2013}. It is not present, however, when N\'eel DWs move due to the spin Hall effect (SHE) generated when a current flows through an adjacent heavy metal layer \cite{Martinez2013}, but in this case the DW is tilted towards Bloch configuration as the current density increases, which reduces the efficiency of the SHE and leads to a saturation in the maximum velocity achievable. It is also absent in systems that exhibit antiferromagnetic coupling, such as antiferromagnets \cite{Shiino2016}, ferrimagnets at angular momentum compensation \cite{Siddiqui2018} or synthetic antiferromagnets \cite{Yang2015}, since DW tilting is virtually suppressed due to strong exchange coupling.

Alternative ways to move DWs in perpendicularly magnetized media that do not require external fields nor charge currents are being explored. Some of them are based on using spatially variable physical quantities, such as anisotropy \cite{Li2020,LST_2018,Wen2020,Franken22012} or temperature \cite{Wang2014,Schlickeiser2014}. Although the detailed mechanism that drives the DW in each case is different, both of them are based on the fact that DW energy depends on the spatially variable quantity and, consequently, a force appears that pushes the DW towards regions where its energy is lowered. This force is typically dependent on the local gradient of the spatially variable quantity \cite{Wen2020}, its effect is essentially equivalent to that of an external magnetic field and, as such, WB occurs when the gradient exceeds a certain threshold value.

Here we investigate the possibility of moving N\'eel DWs in perpendicularly magnetized media using an in-plane strain gradient. The main difference with the approaches mentioned before is that now not only DW energy but also its equilibrium orientation depends on in-plane strain. As it will be shown, this leads to substantial changes in DW dynamics with respect to the standard case. In particular, the interplay between the different torques involved keep the internal DW angle bounded preventing the appearance of precessional dynamics inside DW for any arbitrary strain gradient.

\begin{figure}[t]
\centering
\includegraphics[width=8cm, height=7cm]{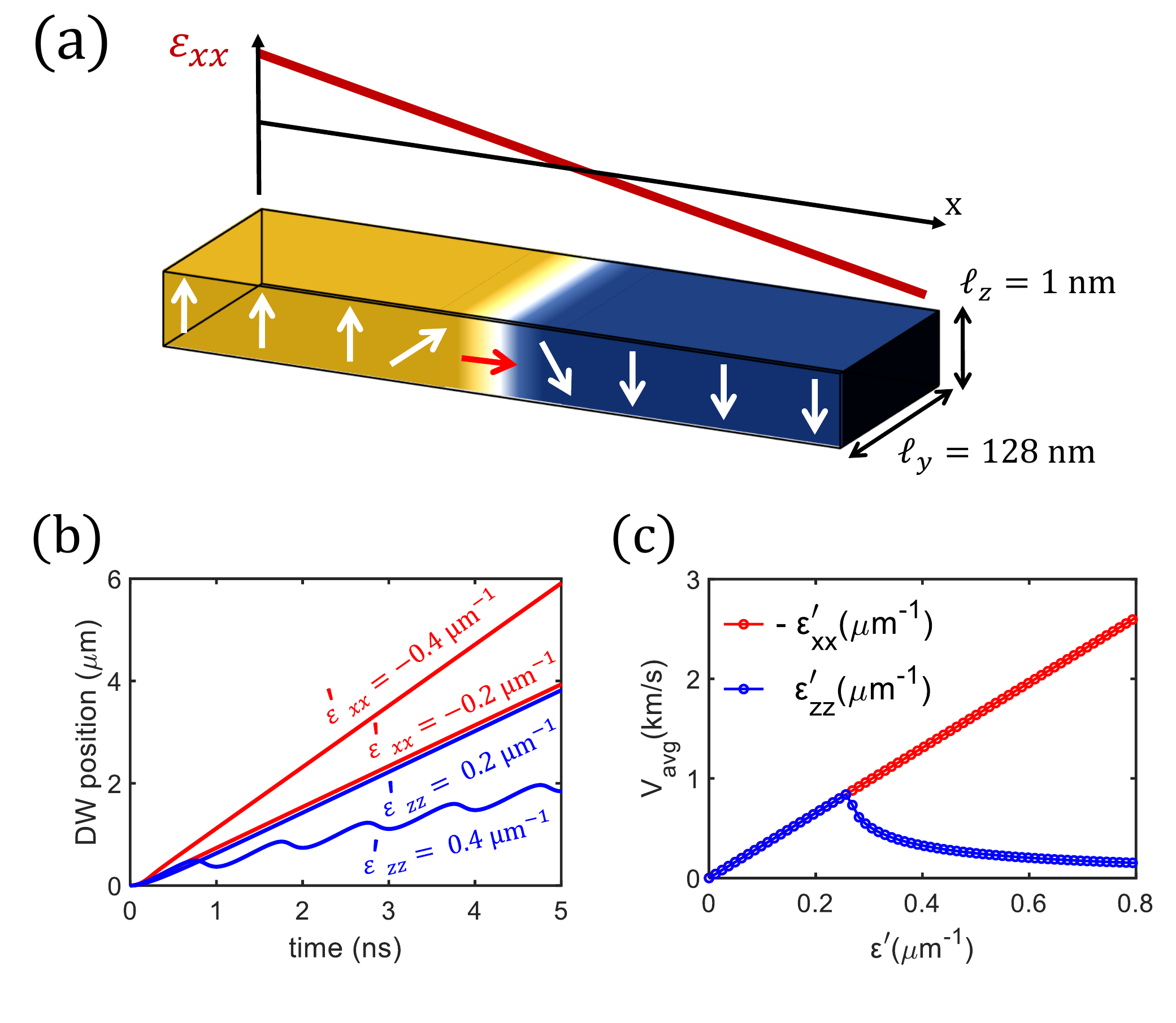}
\caption{ (a) Schematic representation of the system under study. A N\'eel DW is initially located at the center of an infinite ferromagnetic nanostrip subject to an in-plane strain gradient.(b) DW position ($q$) versus time as driven by two different in-plane (blue) and out-of-plane (red) strain gradients. (c) Average DW velocity versus strain slope for the in-plane (blue) and out-of-plane (red) cases.}
\end{figure}


The system under study consists on a N\'eel DW located at the center of an infinite nanowire subject to an in-plane uniaxial strain $\varepsilon_{xx}$ that changes linearly along its length as shown in Fig. 1(a). We assume that this ferromagnetic nanowire is in contact with a heavy metal layer and that the N\'eel configuration is favored over the Bloch one due to the interfacial Dzyaloshinskii-Moriya interaction (DMI) \cite{Martinez2013,Matpar}. Although in the rest of the work we focus on the DW dynamics under strain gradients, we have also explored its static properties under uniform strain (see Sec. I in Supp. Mat. \cite{Supmat}). Uniform strain cannot drive DWs by itself, but its application can tune the DW configuration, and consequently, set on/off the current-driven motion, which is promising for applications.

We will investigate DW dynamics driven by an in-plane strain gradient in absence of any external magnetic field nor charge current. To highlight its peculiarities, we will compare it with the case of a perpendicular strain gradient $\varepsilon_{zz}$. To do so, we firstly use the well-known one-dimensional (1D) model \cite{walker1974,Thiaville_2012,OAyEM2020}, where the magnetization is assumed to change only along $x$. Using the standard ansatz for the DW profile $\theta(x,t)=2\tan^{-1}[\exp({Q\frac{x-q(t)}{\Delta}})]$ and $\varphi(x,t)=\Phi(t)$ where $\theta(x,t)$ and $\varphi(x,t)$ are the spherical coordinates of the magnetization \cite{walker1974,slonczewski1979} the DW energy density reads (See Supp.Mat) \cite{Supmat} 

\begin{equation} \label{DW_energy_strgrad}
\sigma_\text{DW} =\frac{2A}{\Delta}+2\Delta (K_{\text{eff}}+K_{\text{sh}}\sin^2{\Phi}+B_{1}\varepsilon_{xx}^{'}q\cos^2{\Phi})+\pi QD\cos{\Phi} 
\end{equation}

\noindent where $K_{\text{eff}}=K_u -\frac{1}{2}\mu_0 M_s^2$ and $K_{\text{sh}}$ are the effective out-of-plane and shape anisotropy constants. $A$, $D$ and $B_{1}$ are the exchange, DMI and first magnetoelastic constants respectively. Note that $Q=\pm 1$ stands for up-down/down-up DW and $q$, $\Delta$ and $\Phi$ are the DW position, width and angle respectively. It is worth highlighting that the magnetoelastic contribution depends on the DW angle $\Phi$, which is not the case for a perpendicular strain gradient $\varepsilon'_{zz}$ as presented in the supplemental Material \cite{Supmat} . 
Calculating the functional variation of Eq. (\ref{DW_energy_strgrad}) over the DW collective coordinates $(q,\Phi)$ and using LLG equation we obtain the following dynamical equations (see Sec. II in Suppl. Mat. \cite{Supmat} for the derivation of the 1D model Eqs. in the presence of both in-plane and perpendicular strain gradients)

\begin{subequations}
\begin{align}
(1+\alpha^2)\frac{\dot{q}}{\Delta} & = -\alpha \, \Gamma_A(\Phi) + Q\,\Gamma_B(q,\Phi) \label{eqn:q_dot} \\
(1+\alpha^2)\dot{\Phi} & = -Q\,\Gamma_A(\Phi)-\alpha \,\Gamma_B(q,\Phi) \label{eqn:phi_dot}
\end{align}
\end{subequations}

\noindent with 

\begin{subequations}
\begin{align}
\Gamma_A(\Phi) & =  \gamma_{0} \Delta H_\text{mel}\varepsilon'_{xx}\cos^2\Phi \label{eqn:Gamma_A} \\
\Gamma_B(\Phi,q) & =  \gamma_{0} \left[ \left(\frac{H_{\text{sh}}}{2}-H_{\text{mel}}\varepsilon'_{xx}q\right)\sin{2\Phi}-\frac{\pi}{2}QH_{\text{DMI}}\sin{\Phi} \right] \label{eqn:Gamma_B}
\end{align}
\label{eqn:torques}
\end{subequations}

\noindent where $H_{\text{mel}}=\nicefrac{B_{1}}{\mu_{0}M_{s}}$, $H_{\text{sh}}=\nicefrac{2K_{\text{sh}}}{\mu_0 M_s}$ and $H_{\text{D}}=\nicefrac{D}{\mu_{0}M_{s}\Delta}$ are magnetoelastic, shape anisotropy and DMI fields, respectively. In analogy with the pure field-driven case \cite{Mataxass2007,Thiaville_2012}, the term $-\alpha\,\Gamma_A$ in Eq. (\ref{eqn:q_dot}) can be considered as the driving agent that pushes the DW along the direction of decreasing energy, i.e. increasing tensile (compressive) strain if $B_{1}<0$ $(>0)$. On the other hand, the two terms on the RHS of Eq. (\ref{eqn:phi_dot}) can be viewed as the precessional ($-Q\,\Gamma_A$) and damping ($-\alpha\, \Gamma_B$) in-plane torques that govern the internal DW angle dynamics. While the first one is directly linked to the magnetoelastic interaction and only depends on the DW angle [see Eq. (\ref{eqn:Gamma_A})], the second one, which includes shape, DMI and magnetoeslastic contributions, also depends on the DW position [see Eq. (\ref{eqn:Gamma_B})]. This is due to the dependence of the magnetoelastic DW energy on $\Phi$ [see Eq. (\ref{DW_energy_strgrad})] mentioned above, which leads to a term in $\Gamma_B$ that depends on the local strain value at the DW position $(\varepsilon_{xx}^{'}q=\varepsilon_{xx}(q))$. This term has a strong impact on the response of the DW to the strain gradient as we show below.


Fig. 1(b) shows the time evolution of the DW position $q$ for two different values of both in-plane (blue) and perpendicular (red) strain gradients, as predicted from the 1D model. It is worth noting that the dependence of the DW energy on $\varepsilon_{xx}$ and $\varepsilon_{zz}$ is of opposite sign (see Sec II.A Supp.Mat)\cite{Supmat} and, therefore, to produce DW motion in the same direction the sign of the gradient also needs to be opposite. In our case $B_{1}>0$, so $\varepsilon_{xx}^{'}<0$ and $\varepsilon_{zz}^{'}>0$ lead to DW in the positive ($+x$) direction. As can be observed in Fig. 1(b), for the lowest value $(|\varepsilon_{ii}^{'}|=0.2 \, \mu\text{m}^{-^1})$ we find steady motion in both cases with a very similar velocity, whereas for the highest one $(|\varepsilon_{ii}^{'}|=0.4 \, \mu\text{m}^{-^1})$ the response is very different. For perpendicular strain the DW displays turbulent motion with a low average velocity, typical when the system exceeds the WB limit, while for the in-plane case the DW moves steadily with a higher velocity. In Fig. 1(c) we plot the average DW velocity computed over a time window of 5 ns as a function of the strain gradient for both in-plane (blue) and perpendicular (red) cases. For perpendicular strain we observe WB at $\varepsilon_{zz,\text{WB}}^{'}=\frac{\alpha \pi H_{\text{DMI}}}{2H_{\text{mel}}\Delta}\approx 0.26 \, \mu\text{m}^{-^1}$ as one would expect considering that the effect of a perpendicular strain gradient is equivalent to that of an external field $H_\text{eq}=\Delta H_\text{mel}\varepsilon_{zz}^{'}$ (see Sec II.A Supp.Mat)\cite{Supmat} . Below WB the DW velocity is uniform and proportional to the strain gradient $(V=\frac{\gamma_{0}\Delta^{2}}{\alpha}H_\text{mel}\varepsilon_{zz}^{'})$, whereas above it the DW undergoes continuous internal precession, the velocity is no longer uniform and DW mobility is significantly reduced. For the in-plane case, however, no WB is observed and the DW mobility remains constant for arbitrarily high values of the strain gradient. This absence of WB in DW motion driven by an in-plane strain gradient is the main result of our work and, in what follows, we will focus on explaining the mechanism that makes it possible.


\begin{figure}[t]
\centering
\includegraphics[width=13cm, height=10cm]{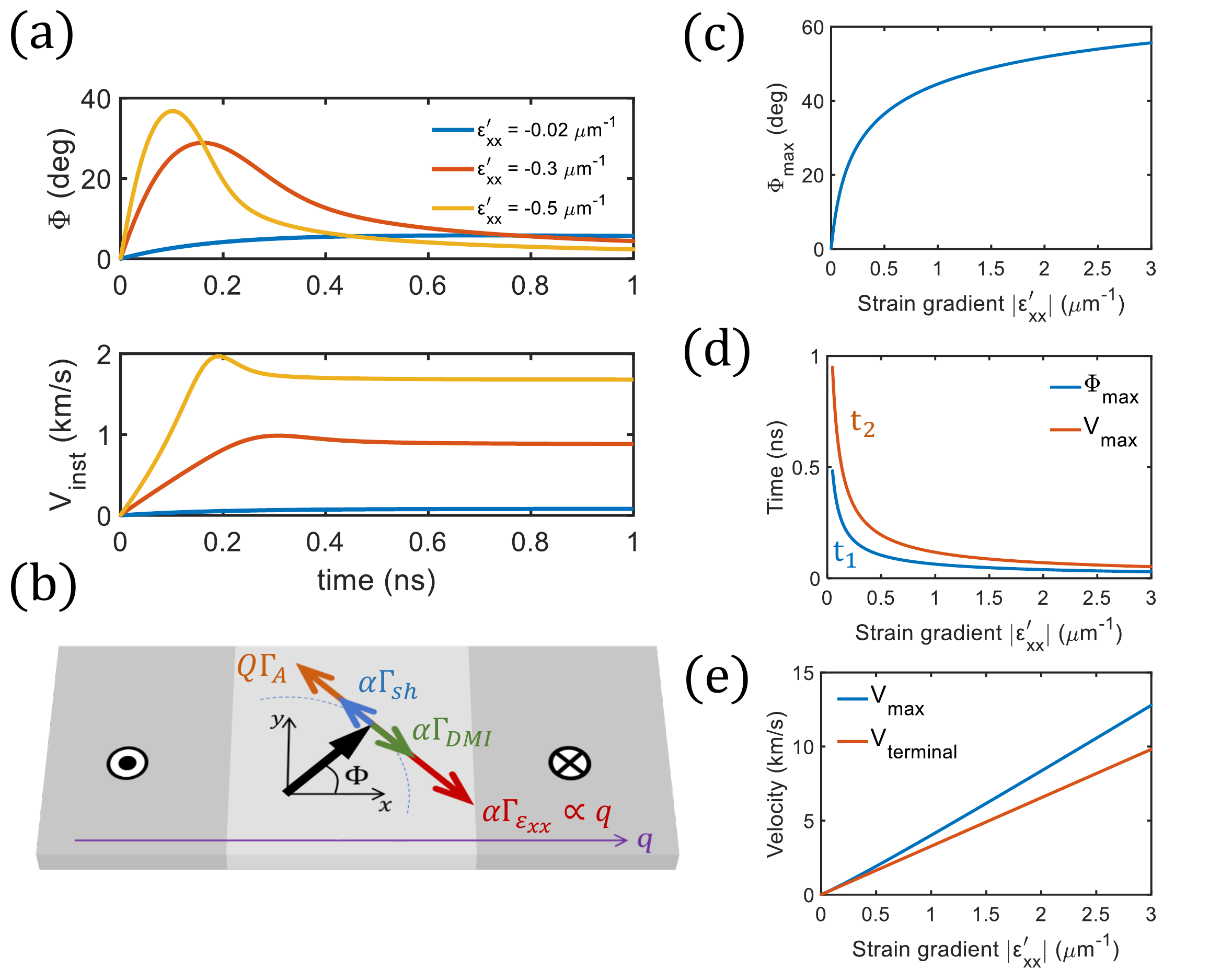}
\caption{(a) DW angle and instantaneous velocity time evolution while driven by an in-plane strain gradient. (b) Schematic representation of the torques contribution on the studied DW dynamics. (c) Maximum tilting of the DW angle versus strain gradient. (d) Characteristic time needed to reach the maximum angle (blue) and maximum velocity (orange). (e) Maximum (blue) and terminal (orange) velocities versus strain gradient.}
\end{figure}

Fig. 2(a) shows the time evolution of the DW angle $\Phi$ (top) and its instantaneous velocity $V_{inst}$ (bottom) for different values of the in-plane strain gradient $\varepsilon_{xx}^{'}$. As can be observed, a transient period, where both $\Phi$ and $V_{inst}$ change non-monotonically, takes place before the DW asymptotically reaches a steady velocity and the angle goes back towards its initial value $(\Phi=0)$. To shed light on this process let us have a look at the different in-plane torques that govern the dynamics of the DW angle in Eq. (\ref{eqn:phi_dot}). On one hand, the precessional component of the driving force, $-Q\Gamma_{A}$, which drives the angle away from its equilibrium orientation. On the other, the damping torque ($-\alpha\Gamma_{B}$), which, in turn, has three terms, i.e. $\alpha\Gamma_\text{sh}=-\alpha \gamma_{0}\frac{H_\text{sh}}{2}\sin{2\Phi}$, $\alpha\Gamma_{\text{DMI}}=\alpha \gamma_{0}\frac{Q\pi H_{\text{DMI}}}{2}\sin{\Phi}$ and $\alpha\Gamma_{\varepsilon_{xx}^{'}}(q)=\alpha \gamma_{0}H_\text{mel}\varepsilon_{xx}^{'}q\sin{2\Phi}$. The first two of them are the standard shape-anisotropy and DMI in-plane damping torques and they are always present, regardless of the nature of the force that moves the DW. The third one, $\alpha\Gamma_{\varepsilon_{xx}^{'}}(q)$, is specific of our case and, unlike the rest of them, its strength depends explicitly on $q$. The four in-plane torques are schematically represented in Fig. 3(b), where their corresponding signs are consistent with an up-down right-handed ($Q=1$) DW moving along the $x>0$ direction.

As soon as the DW starts moving due to the driving torque $-\alpha \Gamma_{A}$, the corresponding precessional component $-Q\Gamma_{A}$ tilts its angle away from the equilibrium orientation and a restoring torque appears trying to bring it back to its orientation at rest. When the driving force is an external field or a perpendicular strain gradient $\varepsilon_{zz}^{'}$, the total restoring torque is given by $\alpha\Gamma_{sh}+\alpha\Gamma_{\text{DMI}}$ and, if the driving force exceeds a certain threshold value, the restoring torque cannot balance the precessional one and WB takes place. However, for an in-plane strain gradient the third term, $\alpha\Gamma_{\varepsilon_{xx}^{'}}(q)$, which opposes the precessional torque, comes also into play, with the peculiarity that its strength increases as the DW moves along the strain gradient (Fig. 2(b)), therefore contributing to bring the angle closer to its value at rest. In fact, this term guarantees that the DW angle remains bounded during motion regardless of the magnitude of the driving force, since its strength, like that of the precessional torque, is proportional to the strain gradient $\varepsilon_{xx}^{'}$ Eq. (\ref{eqn:torques}.b). Fig.S2 in the Supp.Mat shows the temporal evolution of the torques discussed above while the DW is moving \cite{Supmat}. 

With this idea in mind, the transient behavior observed in Fig. 2(a) can be understood as follows. As the DW starts moving and the angle deviates from equilibrium the strength of the in-plane precessional torque $-Q\Gamma_{A}$ decreases, whereas the in-plane damping torque $-\alpha\Gamma_{B}$ increases. Since they have opposite sign, the total in-plane torque is gradually reduced up to a point where both terms balance out $(Q\Gamma_{A}+\alpha\Gamma_{B}=0)$ and the tilting angle reaches its maximum deviation $(\Phi_{max})$. As shown in Fig. 2(c), this maximum deviation $\Phi_{max}$ increases with the strain gradient $\varepsilon_{xx}^{'}$ but it saturates around $\frac{\pi}{3}$, proving our finding that the angle remains bounded no matter how large the strain gradient is. After $\Phi_{max}$ is reached at a certain time $t_{1}$ the total in-plane torque changes sign $(|\alpha\Gamma_{B}| >|Q\Gamma_{A}|)$ and $\Phi$ starts decreasing, whereas the DW velocity continues increasing and it reaches its maximum value $V_{max}$ at a later time $t_{2}$. Fig. 2(d) shows how the two characteristic times of this transient dynamics, $t_{1}$ and $t_{2}$, depend on the strain gradient. As can be observed, they decrease in a similar fashion. As the DW moves further into the region of increasing strain, the term $\alpha\Gamma_{\varepsilon_{xx}^{'}}$ becomes dominant $(|\Gamma_{\varepsilon_{xx}^{'}}| \gg |\Gamma_{DMI}| ,|\Gamma_{sh}|)$ and the DW angle gradually goes back to its initial N\'eel orientation according to $\Phi(q)=\tan^{-1}\left(\frac{Q\Delta}{2\alpha q}\right )$ (see Sec II B Supp.Mat)\cite{Supmat} . The DW velocity, meanwhile, asymptotically reaches a terminal value given by

\begin{equation}
\label{eq:v_term}
v_\text{term} =\frac{\gamma_{0}H_{mel}\Delta^{2}}{\alpha}\varepsilon_{xx}^{'}
\end{equation}

\noindent Fig. 2(e) shows both the terminal and maximum velocities as a function of the strain gradient. While the first one displays a perfectly linear trend, the second one has a slightly stronger dependence so that the difference between them increase with the strain gradient.


\begin{figure}[t]
\centering
\includegraphics[width=7cm, height=10cm]{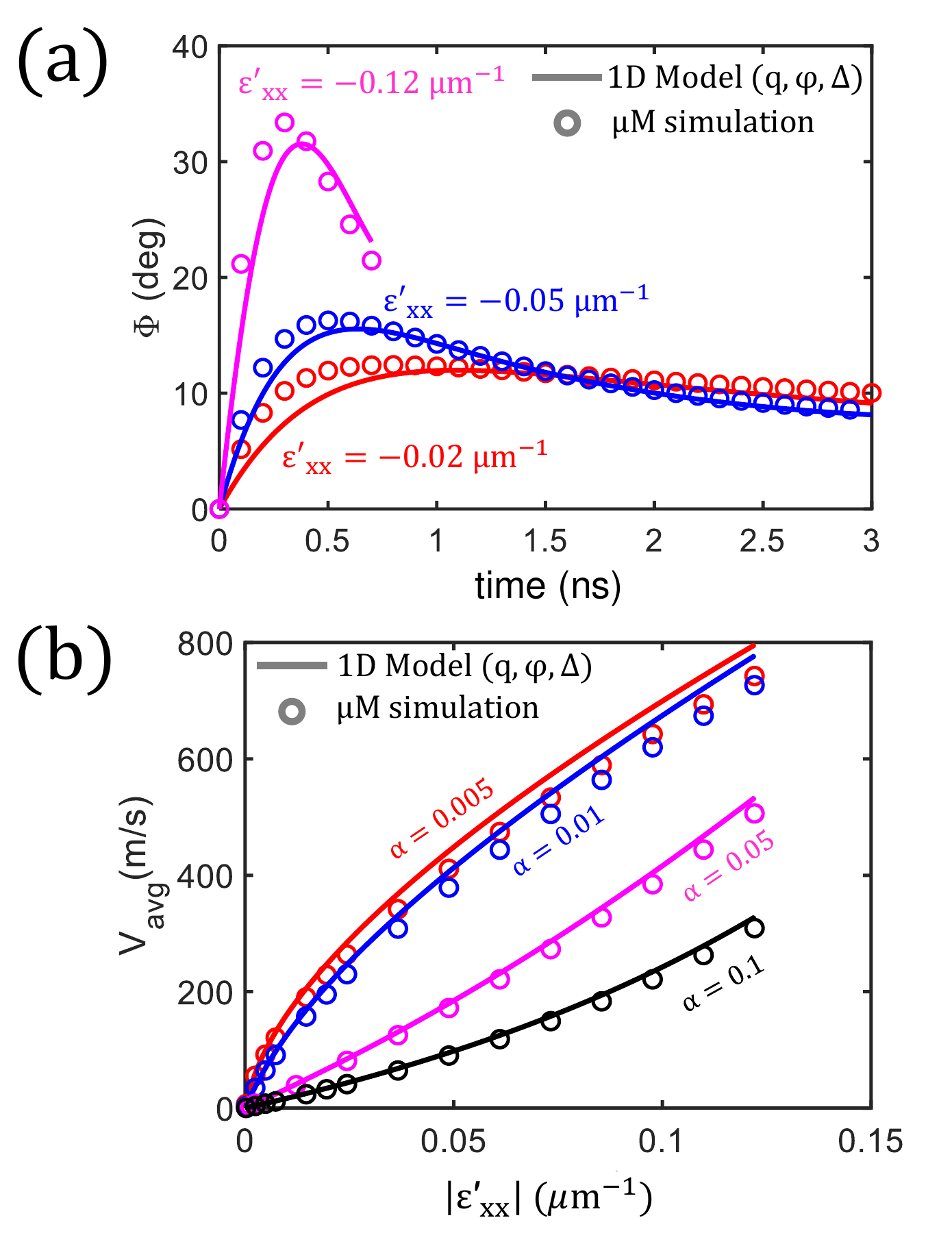}
\caption{(a) Time evolution of the DW angle $\Phi$ for different values of the strain gradient obtained from micromagnetic simulations (dots) and the 1D model (lines). (b) Average DW velocity as a function of the strain gradient for different values of the damping constant $\alpha$ as obtained from micromagnetic simulations (dots) and the 1D model (lines).}
\end{figure}

In practice, there are limitations to the magnitude of the strain gradient that can be applied and the distance over which it can be maintained. Not only a large strain gradient can introduce mechanical damage in the device but also, if strain is large enough to compete with perpendicular anisotropy ($|B_{1}\varepsilon_{xx}|\sim K_\text{eff}$), which in our case happens when $\varepsilon_{xx}\sim 0.06$, the nucleation on in-plane domains start taking place and the system depicted in Fig. 1(a), with a DW separating two antiparallel domains, is no longer stable.

To explore the plausibility of our proposal under realistic conditions micromagnetic simulations were performed using MuMax3 \cite{mumax3} with the same material parameter values given \cite{Supmat} before and a strain gradient over the full length of the device ($1\,\mu\text{m}$). Fig. 3(a) shows the time evolution of the DW angle $\Phi$ for different values of $\varepsilon'_{xx}$ as computed from both micromagnetic simulations (dots) and the 1D model (lines). As can be observed, the simulations support the results shown in Fig. 2, namely that the internal angle remains bounded even for the highest values of the strain gradient. To get a good agreement between the 1D model and the simulations it was necessary to take into account in the former the sizable variations of the DW width \cite{Mougin2007} as it goes deeper into the highly strained region and its effect on the demagnetizing factors (see Supp.Mat) \cite{Supmat,Skaugen2019}.   

Fig. 3(b) shows the average DW velocity $v_\text{avg}$ as a function of the strain gradient for different values of $\alpha$ obtained from micromagnetic simulations (dots) and from the 1D model (lines). Due to the limitations mentioned before, the terminal velocity predicted by the 1D model Eq.(\ref{eq:v_term}) is not accessible in our system. On the other hand, the non-linear dependence with $\varepsilon'_{xx}$ observed for all values of $\alpha$ is due to the fact that the velocity is averaged over a short time interval in which the DW velocity is highly non-uniform, unlike in Fig. 1(c), where the time interval ($5\,\text{ns}$) is significantly larger than this transient period so that $v_\text{avg}\approx v_\text{term}$. In any case, Fig. 3(b) confirms that the absence of WB revealed from the 1D model remains true (see Sec II.C Supp.Mat)\cite{Supmat} . Moreover, our simulations show that large DW velocities can be achieved using in-plane strain gradients under realistic conditions, especially if we take into account that, as shown in our previous publications \cite{APL_skyrmions_strain_gradient,PRAppl_no_SHE}, in-plane strain gradients in the order of $10^{-2} \,\mu\text{m}^{-1}$ can be easily realized in hybrid ferromagnetic/piezoelectric devices by applying moderate voltages between conveniently located electrodes over the piezoelectric substrate.


In conclusion, our theoretical study shows that DW dynamics driven by an in-plane strain gradient in perpendicularly magnetized systems is qualitatively different from the response to other driving forces, such as external field, spin-polarized current or perpendicular strain gradients. In particular, no Walker breakdown takes place in our system and, ideally, a DW velocity proportional to the strain gradient is obtained regardless of its magnitude. We show that the origin of this phenomenon lies in the fact that the DW internal angle in equilibrium depends on strain, which leads to a dynamic torque that opposes tilting and whose strength increases as the DW moves towards increasingly strained regions, therefore preventing the onset of internal DW oscillations. On the other hand, the maximum DW velocity achievable with our approach is not limited by their intrinsic dynamic properties but on the feasibility of keeping perpendicularly magnetized domains stable in regions of high in-plane strain. In any case, our micromagnetic simulations show that averaged velocities in the order of $500 \, \text{m/s}$ can be achieved under realistic conditions. These velocities are in the range of SOTs driven DW motion however with much less energy dissipation since Joule heating is absent. Furthermore, our system reveals complex dynamics where DW angle and, therefore, also its inertia, are not uniquely determined by its velocity but they can largely be tuned with strain, which opens a new avenue to explore dynamic phenomena. 

\begin{acknowledgements}
	{We gratefully acknowledge financial support from the European Union H2020 Program under (MSCA MagnEFi ITN Grant Number 860060), from the Ministerio de Education y Ciencia through the project (MAT2017-87072-C4-1-P), from the Ministerio de Ciencia e Innovacion under the project (PID2020-117024GB-C41) and from Consejeria de Educaci\'on of Castilla y Le\'on under the projects (SA114P20) and (SA299P18).}
\end{acknowledgements}

\bibliographystyle{apsrev4-1}
\bibliography{references}

\begin{thebibliography}{31}
\bibitem [{Mat()}]{Matpar}%
  \BibitemOpen
  \href@noop {} {}\bibinfo {note} {Typical material parameter values to
  stabilize N\'eel domain walls with right-handed chirality were chosen:
  $M_{s}=0.58 \, \text{MA}\,\text{m}^{-1}$, $A=30 \, \text{pJ}\,
  \text{m}^{-1}$, $D=-1.8 \, \text{mJ}/\text{m}^2$, $K_{u}=0.9 \, \text{MJ} \,
  \text{m}^{-3}$, $B_{1}=1.2 \, \text{MJ} \, \text{m}^{-3}$ ($\lambda_{s}=-5
  \times 10^{-5}$, $C_{11}=298 \, \text{GPa}$, $C_{12}=133 \, \text{GPa}$) and
  $\alpha=0.05$.}\BibitemShut {Stop}%
\end{thebibliography}


\end{document}